\documentclass[journal]{IEEEtran}
\ifCLASSINFOpdf
\else
\fi
\usepackage{amsmath, amsthm, amssymb}
\usepackage{algpseudocode}
\usepackage{algorithm}
\usepackage{multirow}
\usepackage{array}
\usepackage[lofdepth,lotdepth]{subfig}
\usepackage{booktabs}
\usepackage{bm}
\usepackage{graphicx}
\usepackage{xcolor}
\usepackage{soul}
\usepackage[utf8]{inputenc}
\usepackage[figuresright]{rotating}
\usepackage{subfig}

\usepackage[figuresright]{rotating}
\usepackage{subfig}
\usepackage{float}

\usepackage{cases}
\usepackage{mdwmath}

\usepackage{tcolorbox}
\usepackage{authblk}
\usepackage{colortbl}
\usepackage[normalem]{ulem}
\usepackage{siunitx}
\usepackage{hyperref}
\usepackage{caption}
\usepackage{graphicx}

\begin{document}
 
\newtheorem{theorem}{Theorem}
\newtheorem{lem}{Lemma}
\newtheorem{cor}{Corollary}

\theoremstyle{definition}
\newtheorem{defn}[theorem]{Definition} 
%
\title{Convolutional Sparse Support Estimator Based Covid-19 Recognition from X-ray Images}
\author[1,3]{Mehmet Yama\c{c}}
\author[1]{Mete Ahishali}
\author[1]{Aysen Degerli}
\author[2]{Serkan Kiranyaz}
\author[2]{Muhammad E. H. Chowdhury}
\author[1]{Moncef Gabbouj}
\affil[1]{Tampere University, Faculty of Information Technology and Communication Sciences, Tampere, Finland}
\affil[2]{Department of Electrical Engineering, Qatar University, Qatar}
\affil[3]{Huawei Technologies Oy (Finland) Co. Ltd, Tampere, Finland}

\maketitle

\begin{abstract}

 Coronavirus disease (Covid-19) has been the main agenda of the whole world since it came in sight in December, 2019. It has already caused thousands of causalities and infected several millions worldwide. Any technological tool that can be provided to healthcare practitioners to save time, effort, and possibly lives has crucial importance. The main tools practitioners currently use to diagnose Covid-19 are Reverse transcription-polymerase chain reaction (RT-PCR) and Computed Tomography (CT), which require significant time, resources and acknowledged experts. X-ray imaging is a common and easily accessible tool that has great potential for Covid-19 diagnosis. In this study, we propose a novel approach for Covid-19 recognition from chest X-ray images. Despite the importance of the problem recent studies in this domain produced not so satisfactory results due to the limited datasets available for training. Recall that Deep Learning techniques can generally provide state-of-the-art performance in many classification tasks when trained properly over large datasets, such data scarcity can be a crucial obstacle when using them for Covid-19 detection. Alternative approaches such as representation-based classification (collaborative or sparse representation) might provide satisfactory performance with limited size datasets, but they generally fall short in performance or speed compared to Machine Learning methods. To address this deficiency, Convolution Support Estimation Network (CSEN) has recently been proposed as a bridge between model-based and Deep Learning approaches by providing a non-iterative real-time mapping from query sample to ideally sparse representation coefficient’ support, which is critical information for class decision in representation based techniques.
 
Main premises of this study can be summarized as follows: (i) a benchmark X-ray dataset, namely QaTa-Cov19, containing over 6200 X-ray images is created. Up to date, this is the largest dataset covering 462 X-ray images from Covid-19 patients along with three other classes; bacterial pneumonia, viral pneumonia, and normal. (ii) In such a scarce and imbalanced dataset the proposed CSEN based classification scheme equipped with feature extraction from a state-of-the-art deep neural network solution for X-ray images, CheXNet, achieves over $98\%$ sensitivity and over $95\%$ specificity for Covid-19 recognition directly from raw X-ray images without any pre- or post-processing. (iii) Having such an elegant Covid-19 assistive diagnosis performance, this study further provides solid evidence that Covid-19 induces a unique pattern in X-rays that can be discriminated with a high accuracy. 
\end{abstract}


%
\IEEEpeerreviewmaketitle

\vspace{-0.5cm}
\section{Introduction}
\label{Introduction}
\IEEEPARstart{C}{oronavirus} disease 2019 (Covid-19) has been declared as a pandemic by the World Health Organization (WHO) two months after its first appearance in December, 2019 in Wuhan, China. It has infected more than 3 million people, caused thousands of causalities and has so far paralyzed the mobility all around the World. The spreading rate of Covid-19 is so high that the number of cases is expected to be doubled every three days if the social distancing is not strictly observed to slow this accretion~\cite{coronaSpreading}. Roughly around half of Covid-19 positive patients exhibit also a comorbidity~\cite{clinical}, making difficult to differentiate Covid-19 from other lung diseases. Automated and accurate Covid-19 diagnosis is critical for both saving lives and preventing its rapid spread in the community. Currently, RT-PCR (Reverse transcription polymerase chain reaction) and CT (computed tomography) are the common diagnosis techniques used today. RT-PCR results are ready at the earliest 24 hours for critical cases and generally take several days to conclude a decision~\cite{CT1}. CT may be an alternative at initial presentation; however, it is expensive and not easily accessible~\cite{erickson1993advanced}. The most common tool that medical experts use for both diagnostic and monitoring the course of the disease is X-ray imaging. Compared to RT-PCR or CT test, having an X-ray image is an extremely low cost and a fast process, usually taking only few seconds. Recently, WHO reported that even RT-PCR may give false results in Covid-19 cases due to several reasons such as poor quality specimen from the patient, inappropriate processing of the specimen, taking the specimen at an early or late stage of the disease~\cite{world2020laboratory}. For this reason, X-ray imaging has a great potential to be an alternative technological tool to be used along with the other tests for an accurate diagnosis.

\begin{figure}[h]
 \centering
  \includegraphics[width=0.9\linewidth]{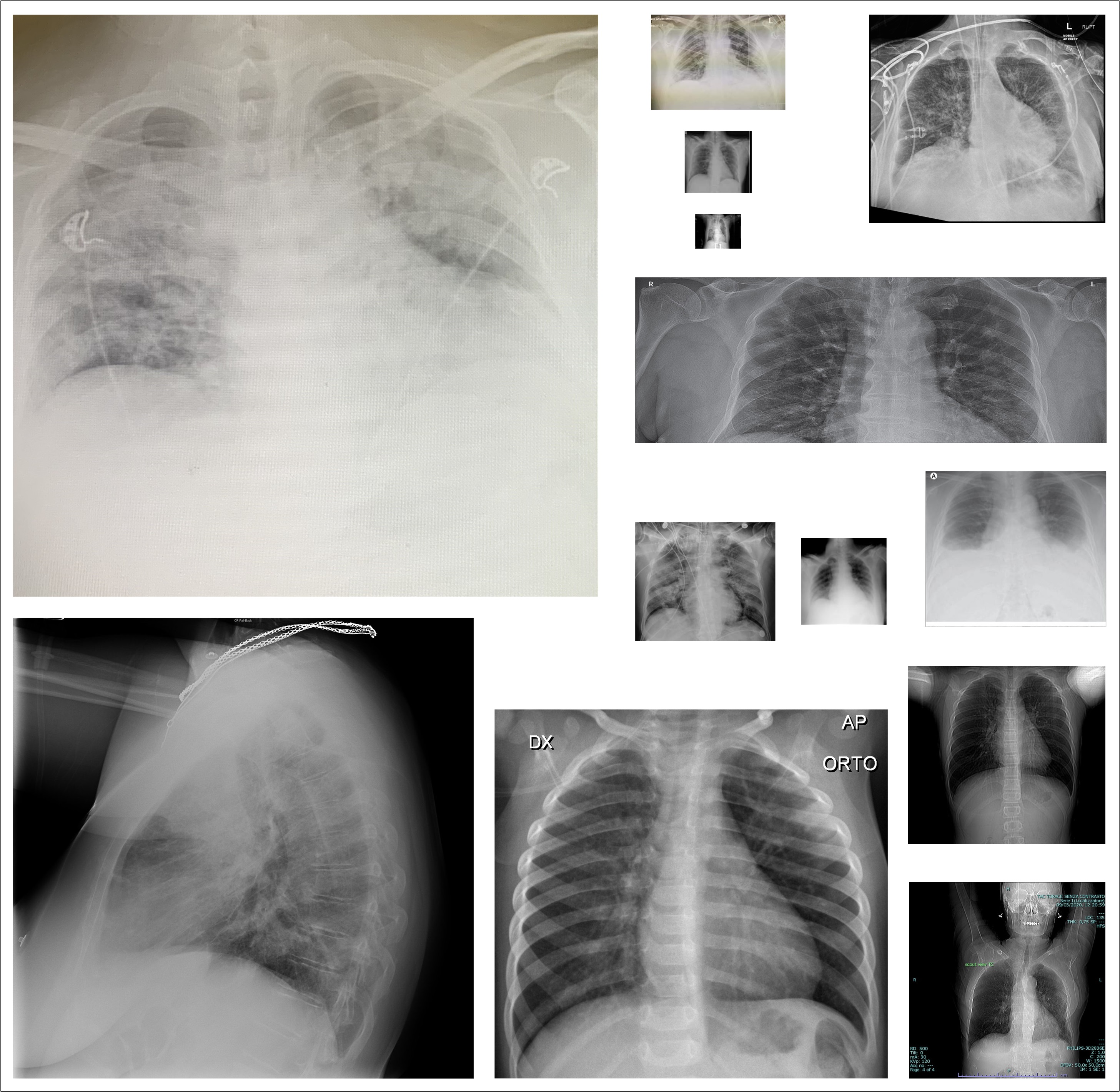}
 \caption{Sample Covid-19 X-ray images from QaTa-Cov19.}
\label{fig:Diversity}
\end{figure}

Accordingly, there are several recent works \cite{Xray1, Xray2, Xray3, exact4} that have been proposed for Covid-19 detection/ classification from X-ray images. However, they use a rather small dataset (the largest containing only a few hundreds of X-ray images), with only a few Covid-19 samples. This makes it difficult to generalize their results in practice. To address this deficiency and provide reliable results, in this study the researchers of \textbf{Qa}tar University and \textbf{Ta}mpere University have compiled the largest \textbf{Cov}id-\textbf{19} dataset, called QaTa-Cov19.  Compared to the earlier benchmark dataset created in this domain, such as  COVID Chestxray Dataset \cite{CovidDataSet1} or Covid-19 DATASET \cite{CovidDataSet2}, QaTa-Cov19 has the followıng unique benchmarking properties. 
First, it is the largest dataset, not only in terms of the number of images (more than 6200 images) but its versatility i.e., QaTa-Cov-19 contains additional major pneumonia categories, such as Viral and Bacterial, along with the control (normal) class. Moreover, this is the most diverse dataset encapsulating X-ray images from several countries (e.g. Italy, Spain, China, etc.) produced by different X-ray machines. Finally, the images are in different quality, resolution and SNR levels as shown in Fig. \ref{fig:Diversity}.

\begin{figure}[h]
	\centering
		\subfloat[]{\includegraphics[width=0.5\textwidth]{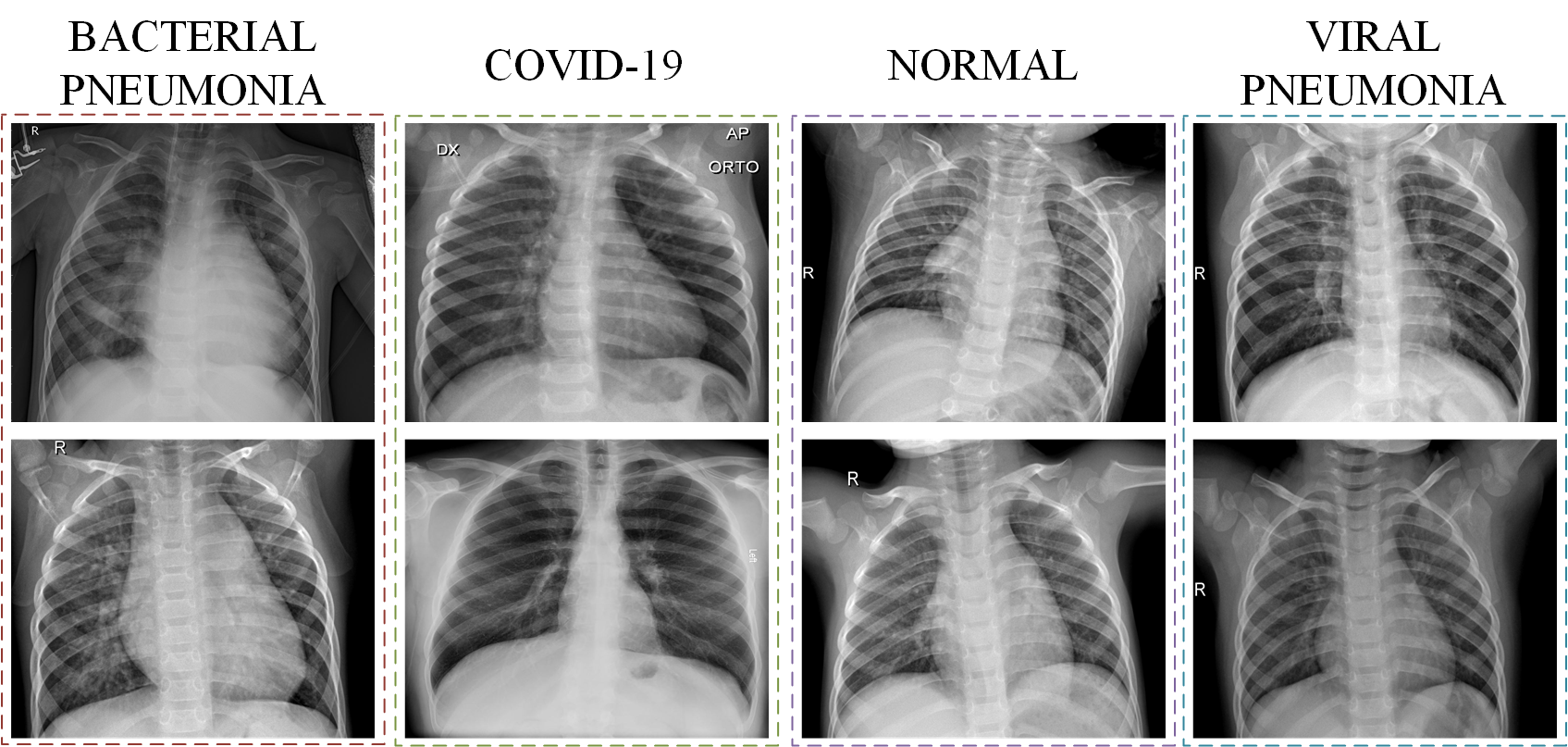}} \\
		\subfloat[]{\includegraphics[width=0.5\textwidth]{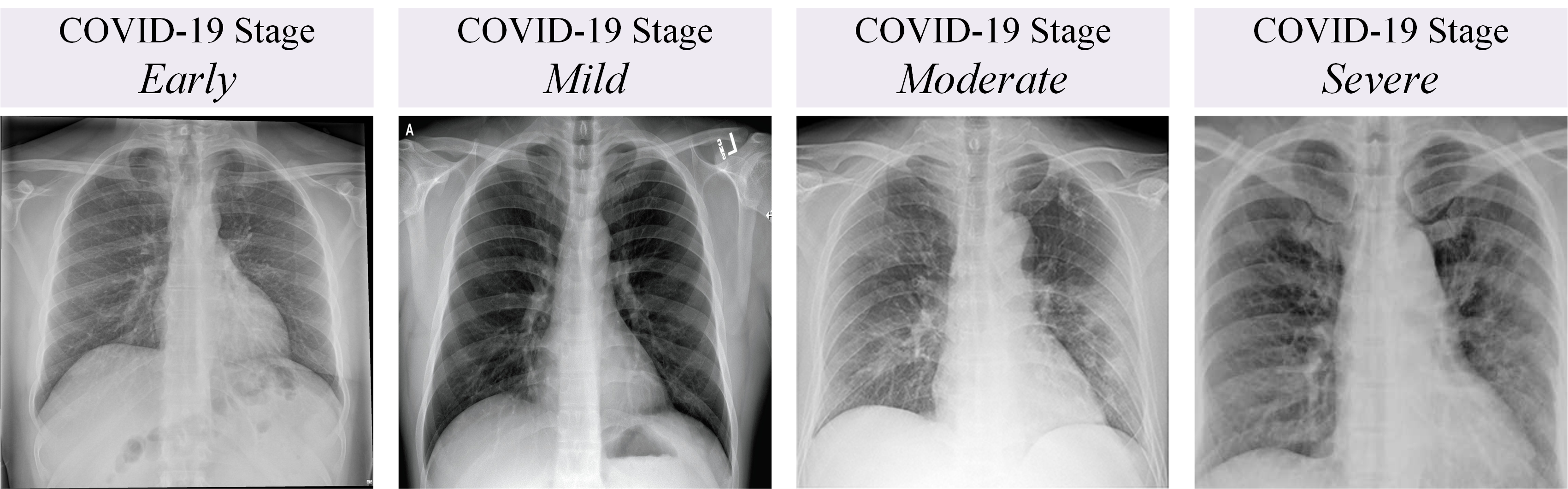}}
	\caption{Sample QaTa-Cov19 X-ray images: (a) X-ray images from different classes. (b) X-ray images from the Covid-19 patients who are in the different stages. } 
	\label{fig:introduction}
\end{figure}

QaTa-Cov19 contains many X-ray images from the Covid-19 patients who are in the early stages; therefore, their X-ray images show mild or no-sign of Covid-19 infestation by the naked eye. Some sample images are shown in Fig.~\ref{fig:introduction}-(b). Another fact which makes the diagnosis far more challenging is that inter-class similarity can be very high for many X-ray images as some samples shown in Fig.~\ref{fig:introduction}-(a). Against such high inter-class similarities and intra-class variations, in this study we aim for a high robustness level. Our primary objective is to achieve the highest sensitivity possible in the diagnosis of Covid-19 induced pneumonia with an acceptable false-alarm rate (e.g. specificity $> 95\%$). In particular, the misdiagnosis of a Covid-19 X-ray image as a normal case should be minimized whilst a small number of false negatives is tolerable. 

In numerous classification tasks, Deep Learning techniques have been shown to achieve state-of-the-art performance in term of both recognition accuracy and their parallelizable computing structures which play an important role especially in real-time applications. Despite their advantages, in order to achieve a desired performance level in a deep model, a proper training over a massive training dataset is usually needed. Nevertheless, this is unfortunately not an option yet for this problem since the available data is still rather limited. 

An alternative supervised approach, which requires a limited number of training samples to achieve satisfactory classification accuracy is representation-based classification \cite{collaborative, SRC1, SRC2}. In representation-based classification systems, a dictionary, whose columns consist of the training samples that are stacked in such a way that a subset of them corresponding to a class, is pre-defined. A test sample is expected to be a linear combination of all points from the same class as the test sample. Therefore, given a predefined dictionary matrix, $\mathbf{D}$ and a test sample $\mathbf{y}$, we expect the solution $\mathbf{\hat{x}}$ from $\mathbf{y} = \mathbf{D x}$, carry enough information about the class of $\mathbf{y}$. The two well-known representation based classification methodologies are sparse representation-based classification (SRC) \cite{SRC1} and collaborative representation based classification (CRC) \cite{collaborative}. Out of these two, SRC provides slightly improved accuracy by solving a sparse representation problem, i.e., producing a sparse solution $\mathbf{\hat{x}}$ from $\mathbf{y} = \mathbf{D x}$. Then, the location of the non-zero elements of $\mathbf{\hat{x}}$, which is also known as support set, provides us with the class of the query $\mathbf{y}$. Despite improved recognition accuracy, SRC solutions are iterative solutions and can be computational demanding compared to CRC. In a recent work \cite{CSEN}, a compact neural network design that can be considered as a bridge between learning-based and representation-based methodologies was proposed. The so-called Convolutional Support Estimation Network (CSEN) uses a pre-defined dictionary and learns a direct mapping using moderate/low size training set, which maps query samples, $\mathbf{y}$, directly to the support set of representation coefficients, $\mathbf{x}$ (as it should be purely sparse in the ideal case).  

\begin{figure*}[t]
 \centering
  \includegraphics[width=0.99\linewidth]{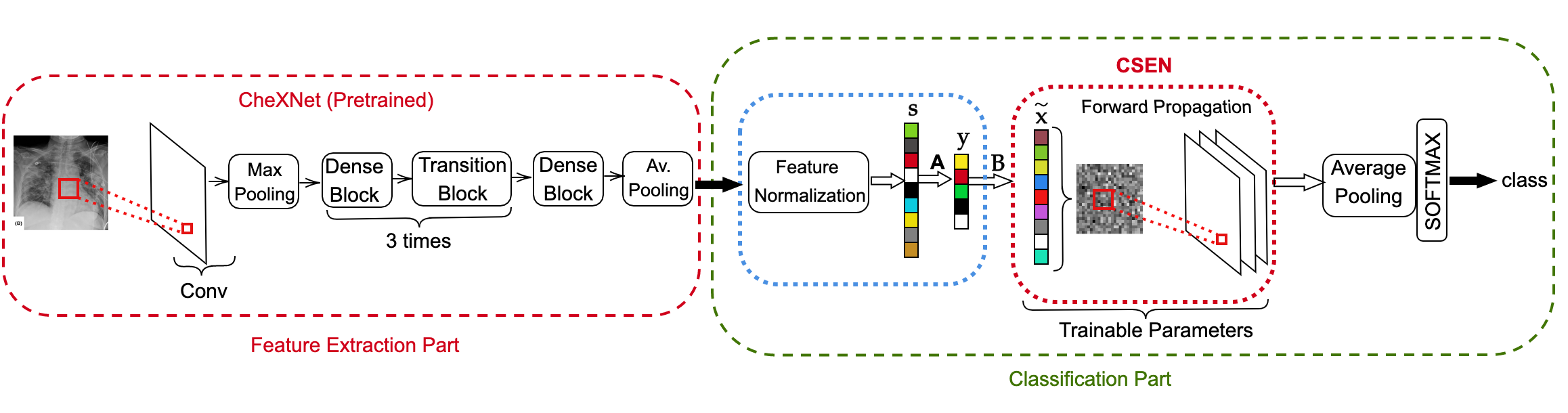}
 \caption{ The proposed approach for Covid recognition from X-ray images. The proposed convolution support estimator network (CSEN) which can be trained from a moderate size training set. The pipeline employs the pre-trained deep neural network for feature extraction. $\mathbf{A}$ is the dimensional reduction (PCA) matrix, the coarse estimation of representation coefficient (sparse in ideal case), $\hat{\mathbf{x}}$ is obtained via the denoiser matrix, $ \mathbf{B} =  \left ( \mathbf{D}^T \mathbf{D} + \lambda \mathbf{I} \right )^{-1} \mathbf{D}^T$, where $\mathbf{D}=\mathbf{A}\mathbf{ \Phi}$ and $\mathbf{ \Phi}$ is the pre-defined dictionary matrix of training samples (before dimensional reduction).}
\label{fig:CSEN general}
\end{figure*}

In this study, to address the aforementioned limitations in Covid-19 diagnosis from X-ray images we propose a CSEN-based approach. Since the largest set of Covid-19 X-ray images ever compiled is used in this study, the proposed approach can be evaluated rigorously against a high-level of diversity to obtain a reliable analysis. The general pipeline of the proposed CSEN based recognition scheme is illustrated in Fig. \ref{fig:CSEN general}. In order to obtain highly discriminative features, we use the recently proposed CheXNet \cite{chexnet}, which is the fine-tuned version of $121$ layer Dense Convolutional Network (DenseNet-121) \cite{DenseNet} by using over $100000$ frontal view X-ray images form $14$ classes. Having the pre-trained CheXNet for feature extraction, we develop two different strategies to obtain the classes of query X-ray images: (i) using collaborative representation-based classification with a proper pre-processing; (ii) a slightly modified version of our recently proposed convolution support estimator (CSEN) models. The proposed CSEN scheme outperforms the competing methods and achieves over $98\%$ of sensitivity and over $95\%$ for specificity in this challenging dataset.

The rest of the paper is organized as follows. In Section~\ref{Notations}, notations and mathematical preliminaries are given with emphasis on sparse representation and sparse support estimation. Then in Section~\ref{Background}, a literature review on deep learning models over X-ray images and representation based classification is presented. The proposed CSEN-based Covid-19 recognition system is introduced in Section~\ref{Proposed Framework} along with two recent alternative approaches that are used as the competing methods. The data collection is also explained in this section. Experimental setup and the main results are provided in Section~\ref{Results}. Finally, Section ~\ref{conclusion} concludes the paper and suggests topics for future research.  

\section{Preliminaries and Mathematical Notations}
\label{Notations}
\subsection{Notations}
In this study, the $\ell_p$-norm of a vector $\mathbf{x} \in \mathbb{R}^n$ is defined as $\left \| \mathbf{x} \right \|_{\ell_p^n} =   \left (  \sum_{i=1}^n \left \vert x_i \right \vert^p \right )^{1/p}$ for $p \geq 1$. On the other hand, the $\ell_0$-norm of the vector $\mathbf{x} \in \mathbb{R}^n$ is defined as $\left \| \mathbf{x} \right \|_{\ell_0^n} = \lim_{p \to 0} \sum_{i=1}^n \left \vert x_i \right \vert^p = \# \{ j: x_j \neq 0 \}$ and the $\ell_{\infty}$-norm is defined as $\left \| \mathbf{x} \right \|_{\ell_{\infty}^n} =  \max_{i=1,...,n} \left (   \left | x_i \right | \right )$. A signal $\mathbf{s}$ is called strictly $k$-sparse if $\left \|  \mathbf{x} \right \|_0 \leq k$. Sparse support set or simply support set, $\Lambda \subset \{1,2,3,...,n \} $
of sparse signal $\mathbf{x}$ can be defined as the set of non-zero coefficients' location, i.e., $\Lambda := \left \{ i:  x_i =0 \right \}$.
\subsection{Sparse Signal Representation}
\label{SSR}
Sparse representation (SR) of a signal $\mathbf{s} \in \mathbb{R}^d $ in a pre-defined set of waveforms, $ \mathbf{\Phi} \in \mathbb{R}^{ d\times n}$, can be defined as representing $\mathbf{s}$ as a linear combination of only a small subset of atoms of in the dictionary $ \mathbf{\Phi}$, i.e, $\mathbf{s} = \mathbf{\Phi }\mathbf{x}$. Defining these sets, which dates back to Fourier's pioneering work \cite{Fourier}, has been excessively studied in the literature. In the early approaches, these sets of waveforms have been selected as a collection of linearly independent and generally orthogonal waveforms (which are called a complete dictionary or basis i.e, $d=n$) such as Fourier Transform, DCT and Wavelet Transform, until the pioneering work of Mallat \cite{mallat1993} on overcomplete dictionaries ($n >>d$). In the last decade, interest in SR research increased tremendously and their wide range of applications includes denoising \cite{denoising}, classification \cite{classification}, anomaly detection \cite{AnomalyDetection, AnomalyDetection2}, Deep Learning \cite{deeplearning} and Compressive Sensing (CS) \cite{CS1,CS2}.

With a possible dimensional reduction that can be satisfied via a compression matrix $\mathbf{A} \in \mathbb{R}^{m \times d} $ ($m << d$), sample can be obtained from $\mathbf{s}$,
\begin{equation}
     \mathbf{y} = \mathbf{A} \mathbf{s} = \mathbf{A}\mathbf{ \Phi} \mathbf{x} =\mathbf{ D} \mathbf{x} \label{CS},
 \end{equation}
where $\mathbf{D} \in \mathbb{R}^{m \times n}$ can be called the equivalent dictionary. Because Eq. \eqref{CS} describes an under-determined system of linear equations, finding the representation coefficient vector $\mathbf{x}$ requires at least one more constraint to have a unique solution. Using the prior information about sparsity, the following representation 
\begin{equation}
\min_\mathbf{x} ~ \left \| \mathbf{x }\right \|_{0}~ \text{subject to}~ \mathbf{D} \mathbf{x} = \mathbf{y} \label{sparse_rep}
\end{equation}
which is also a sparse representation of $\mathbf{x}$ has a unique solution provided that $\mathbf{D}$ satisfies some required properties \cite{spark}. However, the optimization problem in Eq. \eqref{sparse_rep} is a NP-hard. Fortunately, the following relaxation 
\begin{equation}
\min_\mathbf{x} ~ \left \| \mathbf{x }\right \|_{1}~ \text{subject to}~ \mathbf{D} \mathbf{x} = \mathbf{y} \label{l1_rep}
\end{equation}
produces exactly the same solution as that of Eq. \eqref{sparse_rep} provided that $\mathbf{D}$ obeys some criteria \cite{candesRIP} and $m > k (\log (n/k))$. In addition, real world applications generally exhibit not exact sparsity but approximate sparsity. Furthermore, the query sample $\mathbf{y}$ can be corrupted with an additive noise pattern. In this case, the equality constraint in Eq. \eqref{l1_rep} can be further relaxed such as in the Basis Pursuit Denoising (BPDN) \cite{BP}: $\min_{\mathbf{x}} \left \| \mathbf{x} \right \| ~\text{s.t.} ~    \left \| \mathbf{y} - \mathbf{Dx} \right \|  \leq \epsilon$, where $\epsilon$ is a small constant that depends on the noise level. 

We may refer to the Sparse Support Estimation (SE) problem as finding the indices a set, $\Lambda$, of non-zero elements of $\mathbf{x}$~\cite{SE1,SE2}. Indeed, in many applications, SE can be more important than finding the magnitude and sign of $\mathbf{x}$ as well as $\Lambda$, which refers to the sparse Signal Recovery (SSR) via a recovery technique, such as Eq. \eqref{l1_rep}. For example, in a sparse representation based classification system, a query sample $\mathbf{y}$ can be represented with sparse coefficient vector, $\mathbf{x}$, in the dictionary, $\mathbf{D}$ in such a way that when we recover this representation coefficient from $\mathbf{y} = \mathbf{D} \mathbf{x}$, the solution vector $\mathbf{\hat{x}}$ is expected to have a significant number of non-zero coefficients coming from the particular locations corresponding to the class of $\mathbf{y}$.  

Readers are referred to \cite{CSEN} for more detailed literature review on SE and its applications. In the sequel, we briefly summarize the building blocks of the proposed approach.

\section{Background and Prior Art}
\label{Background}
\subsection{CheXNet}
\label{CheXNet}
In the proposed approach, we first use the pre-trained deep network, CheXNet, to extract discriminative features from raw X-ray images. CheXNet was developed for pneumonia detection from the chest X-ray images~\cite{chexnet}. In \cite{chexnet}, it was claimed that their CheXNet can perform even better than expert radiologist in the pneumonia detection problem. This deep neural network design is based on previously proposed DenseNet \cite{DenseNet} that consists of 121 layers. It is first pre-trained over ImageNet dataset \cite{imagenet} and performed transfer learning over $112120$ frontal-view chest X-ray images in the ChestX-ray14 dataset \cite{XrayDataset}.

\subsection{Representation Based Classification}
Given a test sample $\mathbf{y}$, which represents either the extracted features, $\mathbf{s}$, or their dimensionally reduced version, i.e., $\mathbf{y} = \mathbf{A} \mathbf{s}$. In developing the dictionary, training samples are stacked in $\mathbf{D}$ with particular locations in such a way that the optimal support for a given query $\mathbf{y}$ should be the set of all points coming from the same class as $\mathbf{y}$. Therefore, a solution vector, $\mathbf{\hat{x}}$ of $\mathbf{y} = \mathbf{D} \mathbf{x}$ is supposed to have enough information, i.e., the sparse support should be the set of location indices of the training sample from the same class as $\mathbf{y}$. This strategy is generally known as representation-based classification. However, a typical solution $\mathbf{\hat{x}}$ of $\mathbf{y} = \mathbf{D} \mathbf{x}$ is not necessarily a sparse one especially when its size grows with more training samples, which results in a highly under-determined system of linear equations. Fortunately, if one estimates the representation coefficient vector with a sparse recovery design such as $\ell_1$-minimization as in Eq. \eqref{l1_rep}, we can expect that the important non-zero entries of the solution, $\mathbf{\hat{x}}$, are grouped in the particular locations that correspond to the locations of the training samples from the same class as $\mathbf{y}$. This can be a typical example of scenarios where support estimation can be more valuable than the magnitudes and sign recovery as explained in Section~\ref{SSR}.

For instance, \cite{SRC2} proposed a systematic way of determining the identity of face images using $\ell_1$-minimization. The authors develop a three-step classification technique that includes: (i) normalization of all the atoms in $\mathbf{D}$ and $\mathbf{y}$ to have unit $\ell_2$-norm; (ii) estimating the representation coefficient vector via sparse recovery, i.e., $\hat{\mathbf{x}} = \arg \min_{\mathbf{x}} \left \|\mathbf{ x} \right \|_1 \text{s.t} \left \| \mathbf{y} - \mathbf{D} \mathbf{x} \right \|_2 $; and (iii) finding the residuals corresponding to each class via $\mathbf{e_i} = \left \| \mathbf{y} - \mathbf{D_i}  \mathbf{\hat{x}_i} \right \|_2$, where $\mathbf{\hat{x}_i}$ is the group of the estimated coefficients, $\mathbf{\hat{x}}$, that correspond to class $i$.

This technique, which is known as Sparse Representation based Classification (SRC), and its variants have been applied to a wide range of applications in literature \cite{jointsparse, vehicleclassification}, e.g., human action recognition \cite{human-action}, and hyperspecral image classification \cite{hyperspecral}, to name a few. Despite the good recognition accuracy performance of SRC systems, their main drawbacks is the fact that their sparse recovery algorithms (e.g., $\ell_1$-minimization) is iterative methods and computationally costly, rendering them infeasible in real time applications. Later, the authors of \cite{collaborative} introduced Collaborative Representation based Classification (CRC), which is similar to SRC except for the use of traditional $\ell_2$-minimization in the second step; $\mathbf{\hat{x} }= \arg \min_{\mathbf{x}} \left \{ \left \| \mathbf{y }- \mathbf{D}\mathbf{x }\right \|_2^2  + \lambda \left \| \mathbf{x} \right \|_2^2   \right \}$. Thus, CRC does not require an iterative solution to obtain representation coefficient thanks to that $\ell_2$-minimization has a closed form solution, $\hat{\mathbf{x}} = \left ( \mathbf{D}^T  \mathbf{D} + \lambda \mathbf{I}_{n \times n}  \right )^{-1} \mathbf{D}^T  \mathbf{ y}$. Although, the sparsity in $\mathbf{\hat{x}}$ cannot be guaranteed, it has often been reported to achieve a comparable classification performance, especially in small-size training datasets. 

\section{Proposed Approach}
\label{Proposed Framework}

\begin{figure*}[t]
 \centering
  \includegraphics[width=0.95\linewidth]{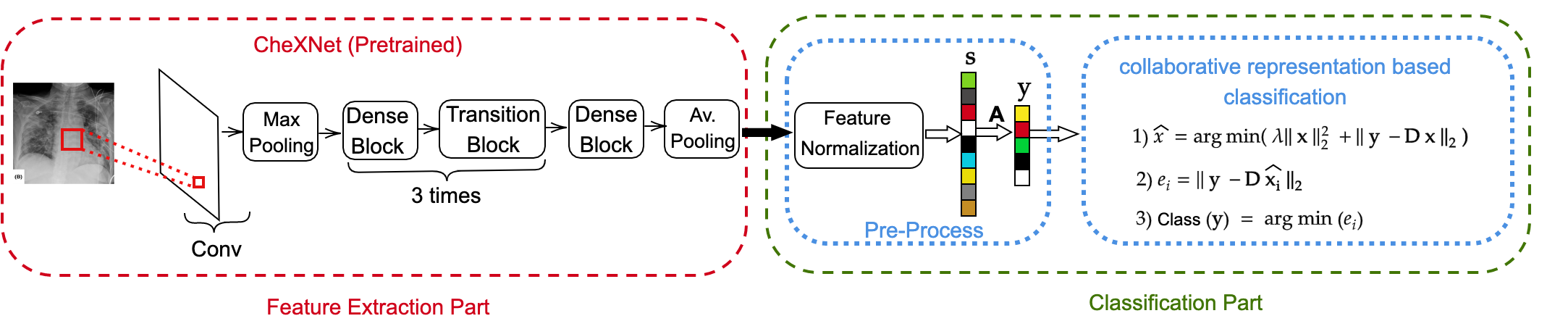}
 \caption{Baseline Approach I: collaborative representation based classification is fed by deep learning based extracted features that are pre-processed. }
\label{fig:CRC general}
\end{figure*}

\begin{figure*}[t]
 \centering
  \includegraphics[width=0.7\linewidth]{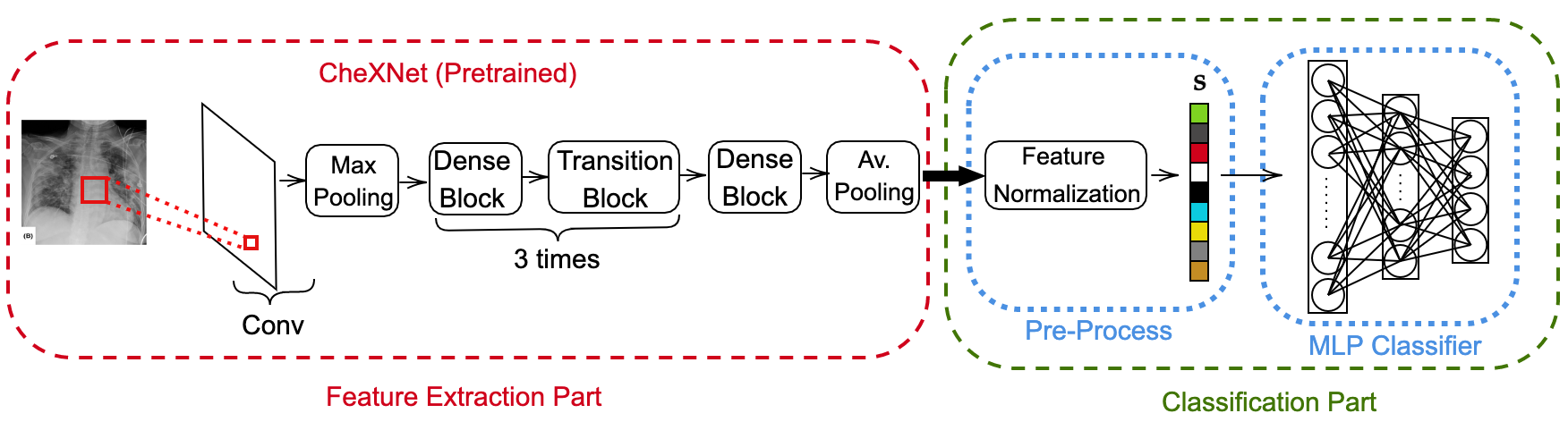}
 \caption{Baseline Approach II: A 5-layer MLP layer is used over the features of CheXNet.}
\label{fig:MLP}
\end{figure*}

\subsection{The Benchmark Dataset: QaTa-Cov19}
\begin{figure}[h!]
 \centering
  \includegraphics[width=0.98\linewidth]{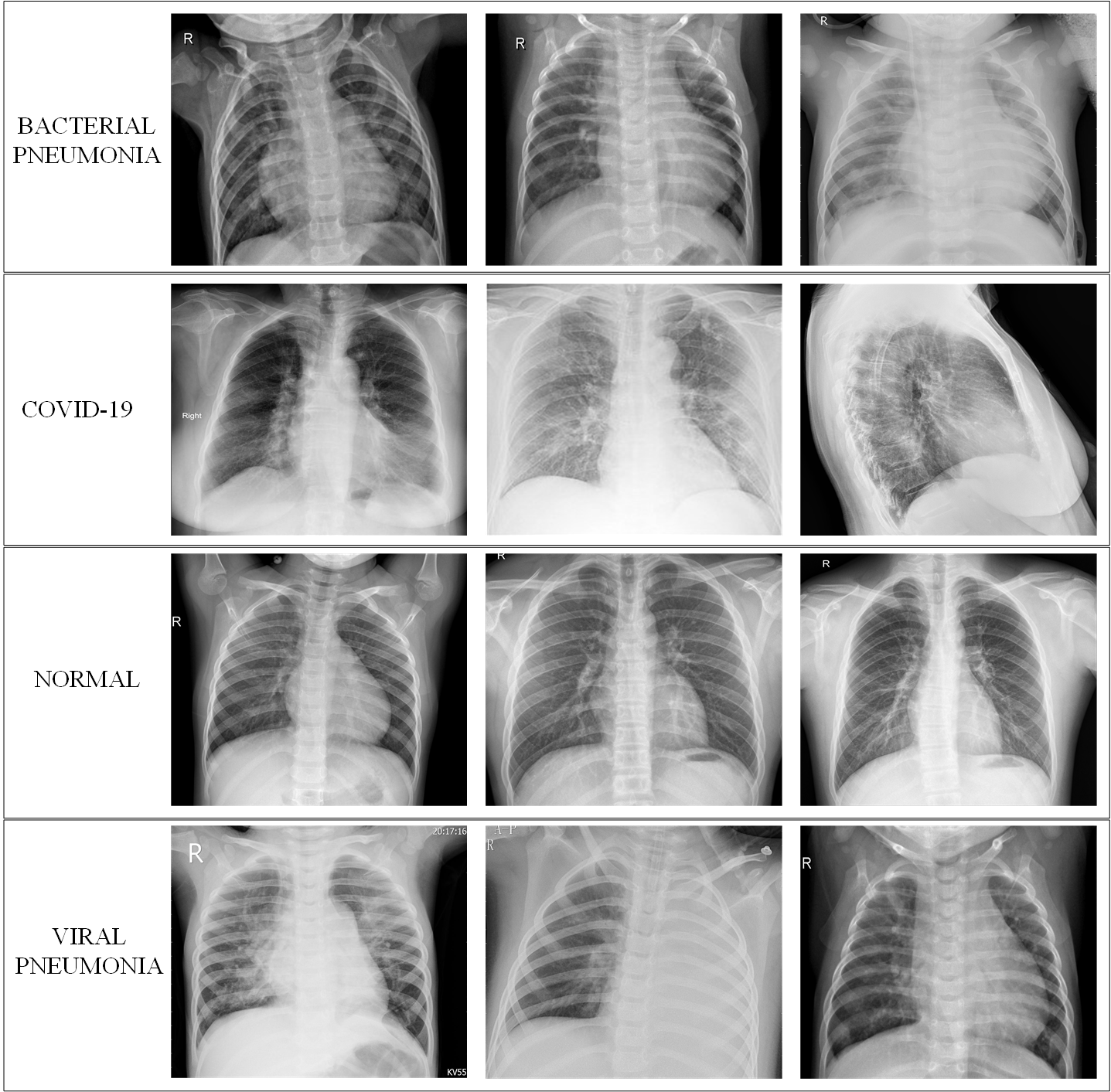}
 \caption{Samples from the benchmark QU-Chest dataset.}
\label{fig:Dataset}
\end{figure}

Covid-19 chest X-ray images were gathered from different publicly available but scattered image sources. However, the major sources of Covid-19 images are Italian Society of Medical and Interventional Radiology (SIRM) COVID-19 Database \cite{CovidDataSet2}, Radiopaedia \cite{CovidDataSet3}, Chest Imaging (Spain) at thread reader \cite{CovidDataSet4} and online articles and news-portals. The authors have carried out the task of collecting and indexing the X-ray images for Covid-19 positive cases reported in the published and preprint articles from China, South Korea, USA, Taiwan, Spain, and Italy, as well as online news-portals (up to 20th April 2020). Therefore, these X-ray images represent different age groups, gender, ethnicity and country. 
Negative Covid19 cases were normal, viral and bacterial pneumonia chest X-ray images and collected from the Kaggle chest X-ray database. Kaggle chest X-ray database contains 5863 chest X-ray images of normal, viral and bacterial pneumonia with varying resolutions \cite{2018chest}. Out of these $5863$ chest X-ray images, $1583$ images are normal images and the remaining are bacterial and viral pneumonia images. Sample X-ray images from QaTa-Cov19 dataset are shown in Fig. \ref{fig:Dataset}.

\subsection{Feature Extraction} 
With their outstanding performance in image classification along with other inference tasks, deep neural networks became a dominant paradigm. However, these techniques usually necessitates a large number of training samples (e.g., several hundred-thousand to millions depending on the network size) to achieve an adequate generalization capability. That is to say, the aforementioned problem of the data scarcity with the Covid-19 case prevents us from training a deep learning technique from scratch. Albeit, we can still leverage their power by finding properly pre-trained models for similar problems. To this end, we use a-state-of-the art pneumonia detection network, CheXNet, whose details are summarized in Section~\ref{CheXNet}. With the pre-trained model, we extract $1024$-long vectors, right after the last average pooling layer. After data normalization (zero mean and unit variance), we obtain a feature vector $\mathbf{s} \in \mathbb{R}^{d=1024}$.

A dimensionality reduction PCA is applied to $\mathbf{s}$ in order to get the query sample, $\mathbf{y} = \mathbf{A} \mathbf{s} \in  \mathbb{R}^m$, where $\mathbf{A} \in \mathbb{R}^{m \times d} $ is PCA matrix ($m<d$).

\subsection{The proposed CSEN-based Classification}

Considering the limited number of training data in our Covid-19 dataset, a representation-based classification can be applied hereafter to obtain the class of $\mathbf{y}$ using the dictionary $\mathbf{\Phi}$ (in the form of $\mathbf{D} = \mathbf{A \Phi}$), whose columns are stacked training samples with class-specific locations.

As discussed earlier, sparse representation-based classification is a support estimation problem which is expected to be an easier task than a sparse signal recovery problem. On the other hand, even if the exact signal recovery is not possible in noisy cases or in cases where $\mathbf{\hat{x}}$ is not exactly but approximately sparse (which is the case in almost all the time in dictionary-based classification problems), it is still possible to recover the support set exactly \cite{exact4,SE1,exact1, Volkan} or partially \cite{Volkan,SE3, partial2}. However, many works in the literature dealing with SE problems tend to first apply a sparse recovery technique on $\mathbf{y}$ to first get $\mathbf{\hat{x}}$, then use simple thresholding over $\mathbf{\hat{x}}$ to obtain a sparse support estimation, $\hat{\Lambda}$. Nevertheless, SSR techniques such as $\ell_1$-minimization are rather slow and their performance varies from one SRR tool to another \cite{CSEN}. In our previous work \cite{CSEN}, we proposed an alternative solution for this handcrafted sparse recovery approach which aims to learn a direct map from test sample $\mathbf{y}$ to the support set $\hat{\Lambda}$. Along with the speed and stability compared to conventional SSR based techniques, and recent deep learning based solutions to SRR problem, CSEN has a crucial advantage of having a compact design that can achieve a good performance level even over scarce training data.

Mathematically speaking, an ideal CSEN is supposed to yield a binary mask $\mathbf{v} \in \left \{ 0,1 \right \}^n$:
\begin{equation}
    {v_i =}
   1    ~ \text{ if $ i \in \Lambda $ }
\end{equation}
which indicates the true support i.e., ${\Lambda} = \left \{  i \in  \left \{  1,2,..,n\right \} : {v}_i =1   \right \}$. In order to approximate this ideal case, a CSEN network, $\mathcal{P} \left ( \mathbf{y}, \mathbf{D} \right )$ produces a probability vector $\mathbf{p} $ which returns a measure about the probability of each index being in ${\Lambda}$ such that $p_i \in \left [ 0,1 \right ]$. Having the estimated probability map, estimating the support can easily be done via $\hat{\Lambda} = \left \{  i \in  \left \{  1,2,..,n\right \} : p_i > \tau   \right \}$, by thresholding $\mathbf{p}$ with $\tau$ where $\tau$ is a fixed threshold.

A CSEN is composed of fully convolutional layers, and as input it takes a proxy, $\mathbf{\tilde{x}}$, of sparse coefficient vector, which is a coarse estimation of $\mathbf{x}$ i.e., $\left ( \mathbf{D}^T \mathbf{D} + \lambda \mathbf{I} \right )^{-1} \mathbf{D}^T\mathbf{y}$ or simply $\mathbf{\tilde{x}=D^Ty}$. Using such a proxy of $\mathbf{x}$, instead of making inference directly on $\mathbf{y}$ has also studied in a few more recent studies. For instance, In \cite{degerli,inference}, the authors proposed reconstruction-free image classification from compressively sensed images. 

\begin{figure}[h]
 \centering
  \includegraphics[width=1\linewidth]{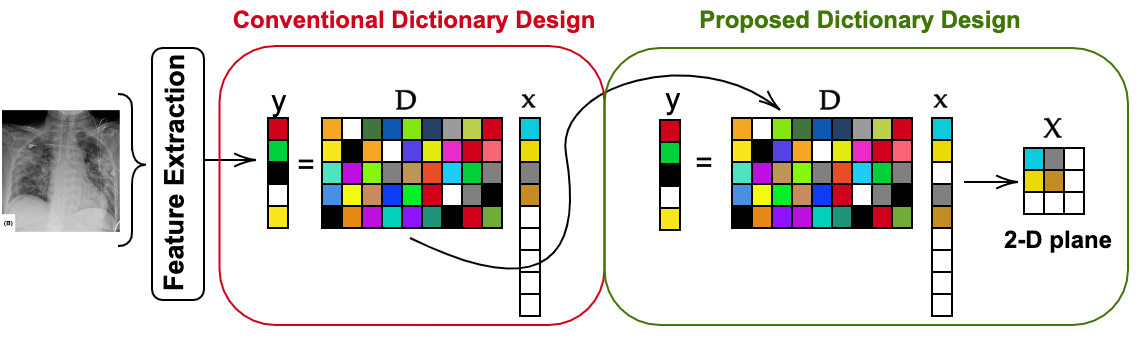}
 \caption{The illustration of proposed dictionary design vs. conventional design in representation based classifiers.}
\label{fig:proposeddic}
\end{figure}

The input vector $\mathbf{\tilde{x}}$ is reshaped to a 2-D plane in order to use is with 2-D convolutional layers. This transformation is performed via re-ordering the indices of the atoms in such a way that the non-zero elements of the representation vector $\mathbf{x}$ for a specific class come together in the 2-D plane. A representative illustration of the proposed dictionary design with compared to the traditional one is shown in Fig.~\ref{fig:proposeddic}. 

Hereafter the proxy $\mathbf{\tilde{x}}$ is convolved with the weight kernels, connecting the input with the next layer with $N$ filters to yield the inputs of the next layer, with the biases $\mathbf{b_1}$ as follows:
\begin{equation}
\mathbf{f_1} = \{S(ReLu(b_1^i + \mathbf{w}_1^i * \tilde{\mathbf{x}}))\}_{i=1}^{N},
\end{equation}
\noindent where $\mathbf{b_1}$ is the weight bias, $S(.)$ is the down- or up-sampling operation and $ReLu(x) = max(0, x)$. In more general form, the $k^{th}$ feature map of layer $l$ is defined as,
\vspace{-0.15cm}
\begin{equation}
    \mathbf{f_l^k} = \textsc{S}(\textsc{ReLu}(b_l^k + \sum_{i=1}^{N_{l-1}}\textsc{conv2D}(\mathbf{w}_l^{ik}, \mathbf{f}_{l-1}^i, '\textsc{ZeroPad}'))).
\end{equation}
Therefore, the trainable parameters of CSEN will be: \\ \noindent $\mathbf{\Theta_{CSEN}}=\big\{ \{\mathbf{w}_1^i, b_1^i\}_{i=1}^{N_1}, \{\mathbf{w}_2^i, b_2^i\}_{i=1}^{N_2}, ... \{\mathbf{w}_L^i, b_L^i\}_{i=1}^{N_L}\big\}$ for a \textit{L} layer CSEN design.

In developing the dictionary that is to be used in the sparse representation based classification, the training samples are stacked-in by grouping of them according to their classes. Thus, instead of using traditional $\ell_1$-minimization formulation as in
Eq. \eqref{l1_rep}, the following group $\ell_1$-minimization formulation may result in increased classification accuracy,
\begin{equation}
    \min_\mathbf{x}  \left \{ \left \|  \mathbf{D}\mathbf{x}-\mathbf{y} \right \|_2^2 + \lambda \sum_{i=1}^{c}\left \|\mathbf{x_{Gi}} \right \|_2  \right \} 
\end{equation}
where $\mathbf{x_{Gi}}$ is the group of coefficients from the $i^th$ class. In this manner, one possible cost function for a SE network would be,
\begin{equation}
E(\mathbf{x}) =  \sum_p (\mathcal{P}_{\Theta}\left (\mathbf{\Tilde{x}} \right )_p- v_p)^2 + \lambda \sum_{i=1}^{c}\left \|\mathcal{P}_{\Theta}\left (\mathbf{\Tilde{x}} \right )_{Gi} \right \|_2. \label{cost}
\end{equation}
where $\mathcal{P}_{\Theta}\left (\mathbf{\Tilde{x}} \right )_p$ is network output at location $p$ and $v_p$ is the ground truth binary mask of the sparse code $\mathbf{x}$. Due to its high computational complexity, we approximate the cost function in \eqref{cost} with a simpler average pooling layer after convolutional layer, which can produce directly the estimated class in our CSEN design. An illustration of proposed CSEN-based Covid-19 recognition is shown in Fig. \ref{fig:CSEN general}.

\subsection{Competing Methods}
This section summarizes the competing methods that are selected among numerous alternatives due to their superior performance levels obtained in similar problems. For a fair comparative evaluations, all classification methods have the same input feature vectors fed to the proposed CSENs.
\subsubsection{Collaborative representation-based classification}
As a possible competing technique to the proposed CSEN based technique which is a hybrid method, CRC \cite{collaborative} is a direct and representation-based classification method. It is a non-iterative support estimation technique, that satisfies faster and comparable classification performance with SRC while it is more stable compared to existing iterative sparse recovery tools as it is shown in \cite{CSEN}. In the first step of CRC, the trade-off parameter of regularized least square solution is set as $\lambda =2*10^{-12}$.

\subsubsection{Multi-layer Perceptron (MLP)  classification}

\begin{figure}[h]
 \centering
  \includegraphics[width=0.65\linewidth]{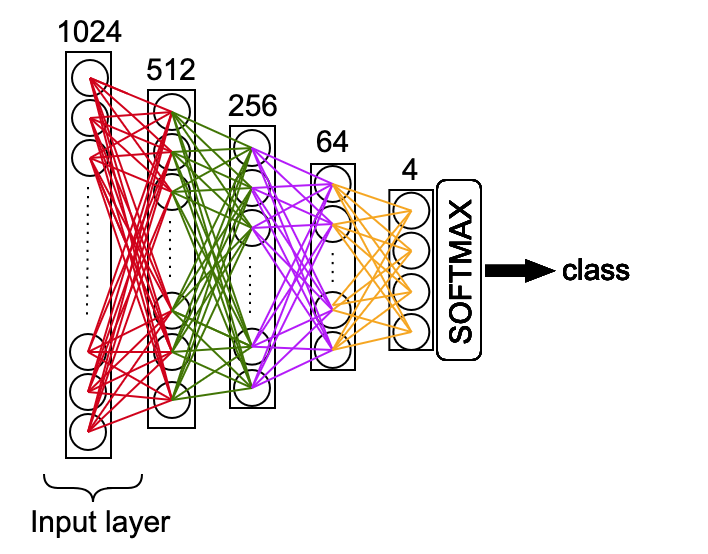}
 \caption{The MLP configuration.}
\label{fig:MLPdesign}
\end{figure}

As one of the most-common classifiers, a 4-hidden layer MLP is used for this problem. For training we used Back-Propagation (BP) with Adam optimization technique \cite{adam}. The network and training hyper-parameters are as follows: learning rate, $\alpha= 10^{-4}$, and moment updates $\beta_1 =0.9$, $\beta_2 =0.999$, and $50$ as the number of epochs. Fig. \ref{fig:MLPdesign} illustrates the network configuration in detail. This network configuration has achieved the best performance among others (deeper and shallower) where deep configurations have suffered from \textit{over-fitting} while the shallow ones exhibit an inferior learning performance. 


\subsubsection{Support Vector Machines (SVMs)}
For a multi-class problem, the first objective is to select the SVM topology for ensemble learning: one-vs-one or one-vs-all. In order to find the optimal topology and the hyper-parameters (e.g. kernel type and its parameters) we first performed a grid-search with the following variations and setting: kernel function \{linear, radial basis function (RBF)\}, box constraint ($C$ parameter) in the range $[1, 10^3]$ with log scale, and kernel scale ($\gamma$ for the RBF kernel) in the range $[10^{-4}, 10^{-2}]$ with log scale.
\subsubsection{k-Nearest-Neighbor (k-NN)}
Finally, we use a traditional approach, k-Nearest Neighbor (k-NN) is used with PCA dimensionality reduction. In a similar fashion, the distance metric and the k-value are optimized by a prior grid-search. The following distance metrics are evaluated: City-block, Chebyshev, correlation, cosine, Euclidean, Hamming, Jaccard, Mahalanobis, Minkowski, standardized Euclidean, and Spearman metrics. The k-value is varied within the range of $[1, 4416]$ with log scale.

\begin{table*}[]
\centering
\caption{Classification Performances of the proposed CSEN and competing methods. 
The best Covid-19 recognition (sensitivity) rates are highlighted.}
\begin{tabular}{|c|c|c|c|c|c|c|c|c|c|c|c|c|}
\hline
         & Bacterial  & Viral & Normal & Covid-19 & Bacterial & Viral & Normal & Covid-19       & Bacterial  & Viral  & Normal  & Covid-19  \\ \hline \hline
         & \multicolumn{4}{c|}{\textbf{Accuracy}} & \multicolumn{4}{c|}{\textbf{Sensitivity}}   & \multicolumn{4}{c|}{\textbf{Specificity}} \\ \hline \hline 
\rowcolor[gray]{.95}NN       & 0.777      & 0.801 & 0.903  & 0.950    & 0.623     & 0.612 & 0.899  & 0.965          & 0.898      & 0.859  & 0.904   & 0.949     \\ \hline
SVM      & 0.771      & 0.788 & 0.928  & 0.928    & 0.586     & 0.632 & 0.911  & 0.981          & 0.916      & 0.837  & 0.933   & 0.924     \\ \hline
\rowcolor[gray]{.95}MLP      & 0.761      & 0.765 & 0.923  & 0.947    & 0.620     & 0.561 & 0.885  & 0.965          & 0.872      & 0.828  & 0.936   & 0.946     \\ \hline
CRC      & 0.820      & 0.827 & 0.928  & 0.955    & 0.758     & 0.550 & 0.922  & 0.968          & 0.869      & 0.913  & 0.930   & 0.954     \\ \hline
\rowcolor[gray]{.95}ReconNET & 0.765      & 0.785 & 0.918  & 0.936    & 0.590     & 0.625 & 0.891  & 0.970          & 0.902      & 0.834  & 0.927   & 0.933     \\ \hline
CSEN1    & 0.793      & 0.805 & 0.926  & 0.955    & 0.656     & 0.642 & 0.906  & \textbf{0.985} & 0.901      & 0.856  & 0.932   & 0.953     \\ \hline
\rowcolor[gray]{.95}CSEN2    & 0.794      & 0.803 & 0.927  & 0.959    & 0.659     & 0.646 & 0.904  & \textbf{0.985} & 0.900      & 0.852  & 0.934   & 0.957     \\ \hline
\end{tabular}
\label{tab:performance}
\end{table*}

\section{Experimental Results}
\label{Results}
\subsection{Experimental Setup}
We have performed our experiments over the QaTa-Cov19 dataset, which consists of normal and three  pneumonia classes: bacterial, viral, and Covid-19. The proposed approach is evaluated using a stratified 5-fold cross-validation (CV) scheme with a ratio of 80\% for training and 20\% for the test (unseen folds) splits, respectively.

\begin{table}[h!]
\centering
\caption{Number of images per class and per-fold before and after data augmentation.}
\begin{tabular}{|c|c|c|c|c|}
\hline
\rowcolor[gray]{.85}Class & \# of Samples & \begin{tabular}[c]{@{}c@{}}Training \\ Samples\end{tabular} & \begin{tabular}[c]{@{}c@{}}Augmented \\ Training Samples\end{tabular} & \begin{tabular}[c]{@{}c@{}}Test \\ Samples\end{tabular} \\ \hline \hline
\begin{tabular}[c]{@{}c@{}}Bacterial\\ Pneumonia\end{tabular} & 2760 & 2208 & 2208 & 552 \\ \hline
\rowcolor[gray]{.95}\begin{tabular}[c]{@{}c@{}}Viral \\ Pneumonia\end{tabular} & 1485 & 1188 & 2208 & 297 \\ \hline
Normal & 1579 & 1263 & 2208 & 316 \\ \hline
\rowcolor[gray]{.95}Covid-19 & 462 & 370 & 2208 & 92 \\ \hline \hline
Total & 6286 & 5029 & 8832 & 1257 \\ \hline
\end{tabular}
\label{numberofsamples}
\end{table} 

Table \ref{numberofsamples} shows the number of X-ray images per class in the QaTa-Cov19 dataset. Since the dataset is unbalanced, we have applied data augmentation to the training set in order to balance the size of each class in the train set. Therefore, the X-ray images in viral and Covid-19  pneumonia, and normal classes are augmented up to the same number as the bacterial pneumonia class in the train set. We use Image Data Generator by Keras to perform data augmentation by applying ZCA whitening with epsilon of $10^{-6}$, randomly rotating the X-ray images in a range of 10 degrees, randomly shifting images both horizontally and vertically within the interval of $[-0.1, +0.1]$. In each CV fold, we use a total of 8832 and 1257  images in the train and test (unseen in the fold) sets, respectively.

The experimental evaluations of SVM, k-NN and CRC are performed using MATLAB version 2019a, running on PC with Intel ® i7-8650U CPU and 32 GB system memory. On the other hand, MLP and CSEN methods are implemented using Tensorflow library \cite{abadi2016tensorflow} with Python on NVidia ® TITAN-X GPU card. For the CSEN training, ADAM optimizer \cite{adam} is used with the proposed default learning parameters: learning rate, $\alpha=10^{-3}$, and moment updates $\beta_1 = 0.9$, $\beta_2 = 0.999$ with only 15 Back-Propagation epochs. Neither grid-search nor any other parameter or configuration optimization was performed for CSEN.

\subsection{Experimental Results}

The same network configurations are used for CSEN as in \cite{CSEN}. Accordingly,
we use two compact CSEN designs: CSEN1 and CSEN2, respectively. The first CSEN network consists of only two hidden convolutional layers, the first layer has 48 neurons and the second has 24. ReLu activation function is used in the hidden layers and the filter size was $3\times3$. On the other hand CSEN2 uses max-pooling and has one additional hidden layer with 24 neurons to perform transposed-convolution. CSEN1 and CSEN2 are compared against the 6 competing methods under the same experimental setup.

For the dictionary construction in $\mathbf{\Phi}$ each CSEN design, $625$ images for each class (from the augmented training samples per fold) are stacked in a such way that the representation coefficient in the 2-D plane, $\mathbf{X}$ has  
$50\times50$ size as shown in Fig. \ref{fig:proposeddic}. The rest of the images in the training set are used to train each CSEN i.e., $1583$ samples from each class. We use PCA dimensional reduction matrix, $\mathbf{A}$ with the compression ratio, $CR  = \frac{m}{d} = 0.5$. Therefore, we have $512 \times 2500$ equivalent dictionary, $\mathbf{D}$, and $2500 \times 512$ denoiser $ \mathbf{B} =  \left ( \mathbf{D}^T \mathbf{D} + \lambda \mathbf{I} \right )^{-1} \mathbf{D}^T$ to obtain a coarse estimation of the representation (sparse in ideal case) coefficients, $\mathbf{\tilde{x}} \in \mathbb{R}^{n = 2500}$. Hereafter, the CSEN networks are trained to have class of $\mathbf{y}$ from input $\mathbf{\tilde{x}}$ as illustrated in Fig., \ref{fig:CSEN general}. 

Due to the lack of other learning-based SE studies in the literature, we chose a deeper network compared to CSEN designs to investigate the role of network depth in this problem. ReconNet \cite{reconnet} was proposed as a non-iterative deep learning solution to compressive sensing problem i.e., $ \mathbf{ \hat{s} } \leftarrow    \mathcal{P} \left ( \mathbf{y} \right ) $ and it is one of the state-of-the-art in compressively sensed image recognition task. It consists of 6 fully convolutional layers and one dense layer in front of the convolutional ones, which act as the learned denoiser for the mapping from $\mathbf{y} \in \mathbb{R}^m$ to $\mathbf{\tilde{s}} \in \mathbb{R}^d$. Then, the convolutional layers are responsible for producing the reconstructed signal, $\mathbf{\hat{s}}$ from $\mathbf{\tilde{s}}$. Therefore, by replacing this dense layer with the denoiser matrix $\mathbf{B}$, this network can be used as a competing method.

Both CSEN and the modified ReconNet use $\mathbf{\tilde{x}}$ as a input, which is produced using an equivalent dictionary $\mathbf{D}$ and its pseudo-inverse matrix $\mathbf{B}$. 


\begin{figure*}[ht!]
 \centering
  \includegraphics[width=1\linewidth]{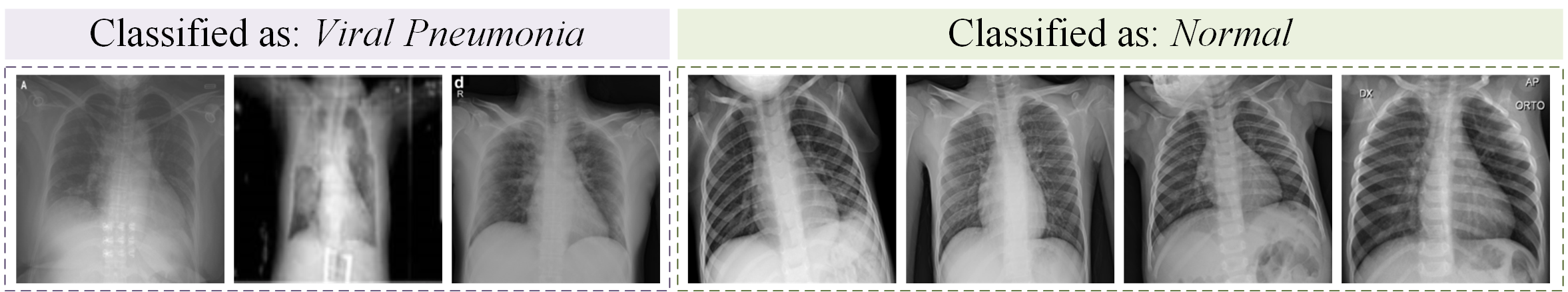}
  \caption{False Negatives of proposed Covid-19 recognition scheme.}
\label{fig:FN}
\end{figure*}

In designing the dictionary of CRC system, all training samples are stacked in the dictionary, $\mathbf{\Phi}$, i.e., 2208 samples from each class. The same PCA matrix used in CSEN based recognition, $\mathbf{A}$ is applied to features, $\mathbf{s} \in \mathbb{R}^{d=1024}$. Therefore, a dictionary $\mathbf{D}$ of size $512 \times 8832$ and the corresponding denoiser matrix $\mathbf{B}$ of size $8832 \times 512$ are used in the CRC framework. 

\begin{table}[h]
\begin{center}
\caption{The number of network parameters of each method. }
\begin{tabular}{@{}|l|l|l|l|l|@{}}
\toprule
                                                                     & MLP     & CSEN1  & CSEN2  & ReconNet \\ \midrule
\begin{tabular}[c]{@{}l@{}}\# of trainable\\ parameters\end{tabular} & 672,836 & 11,089 & 16,297 & 22,914   \\ \bottomrule
\end{tabular}
\label{tab:comparison compactness}
\end{center}
\end{table}

\begin{table}[h!] \scriptsize
\caption{Computation times (sec) of each method over 1257 test images.}
\begin{tabular}{@{}lllllll@{}}
\toprule
                                                                      & \multicolumn{1}{c}{\begin{tabular}[c]{@{}c@{}}CRC \\ (light)\end{tabular}} & CRC     & CSEN1   & CSEN2   & ReconNet & MLP     \\ \midrule
\begin{tabular}[c]{@{}l@{}}Computation \\ Time (in sec.)\end{tabular} & 13.4176                                                                    & 40.7878 & 0.2196 & 0.2272 & 0.2993  & 0.2935 \\ \bottomrule
\end{tabular}
\label{tab:comparison}
\end{table}

The classification performance of the proposed CSEN-based approach and the competing methods is presented in Table \ref{tab:performance}. As can be easily observed from the Table \ref{tab:performance}, the proposed approaches surpass all competing methods in Covid-19 recognition performance by achieving $98.5\%$ sensitivity, and over $95\%$ specificity. As shown in Table \ref{tab:comparison compactness}, compared to MLP and ReconNet, the proposed CSEN designs are very compact, and computationally efficient. This is evident in Table \ref{tab:comparison} where the computational complexity (measured as total computation, time over the 1257 test images) is reported. 

When compared against CRC in particular, CSEN-based classification has two advantages; computational efficiency and, a superior Covid-19 recognition performance. The computational efficiency comes from the fact that a larger size dictionary matrix (of size of $512 \times 8832$) is used in CRC and hence, this requires more computations in terms of matrix-vector multiplications. Furthermore, saving the trainable parameters ($\sim  16k$) and a light dictionary matrix coefficients ($\sim  1280k$) in the test device is more memory efficient compared to saving coefficients ($   \sim  4521k$) of larger size dictionary used in CRC.


\begin{table}[h!]
\begin{center}
\caption{Performance of CRC algorithm when the dictionary (size of 625 per class) that is used in CSEN is used. }
\begin{tabular}{|
>{\columncolor[HTML]{FFFFFF}}l |
>{\columncolor[HTML]{FFFFFF}}l |
>{\columncolor[HTML]{FFFFFF}}l |
>{\columncolor[HTML]{FFFFFF}}l |}
\hline
{\color[HTML]{000000} \textbf{}} & \multicolumn{3}{c|}{\cellcolor[HTML]{FFFFFF}{\color[HTML]{000000} \textbf{CRC (Light)}}}                                             \\ \hline
{\color[HTML]{000000} }          & {\color[HTML]{000000} \textbf{Accuracy}} & {\color[HTML]{000000} \textbf{Sensitivity}} & {\color[HTML]{000000} \textbf{Specificity}} \\ \hline
{\color[HTML]{000000} Bacterial} & {\color[HTML]{000000} 0.8129}            & {\color[HTML]{000000} 0.7464}               & {\color[HTML]{000000} 0.8650}               \\ \hline
{\color[HTML]{000000} Viral}     & {\color[HTML]{000000} 0.8163}            & {\color[HTML]{000000} 0.5461}               & {\color[HTML]{000000} 0.8998}               \\ \hline
{\color[HTML]{000000} Normal}    & {\color[HTML]{000000} 0.9267}            & {\color[HTML]{000000} 0.9170}               & {\color[HTML]{000000} 0.9299}               \\ \hline
{\color[HTML]{000000} Covid-19}  & {\color[HTML]{000000} 0.9564}            & {\color[HTML]{000000} 0.9394}               & {\color[HTML]{000000} 0.9578}               \\ \hline
\end{tabular}
\label{tab:crc light}
\end{center}
\end{table}

For further analysis, we also tested the CRC framework by using the light dictionary (of size $512 \times 2500$) used in CSEN based recognition.
We called it CRC (light), and as it can be seen in Table \ref{tab:crc light}, the performance of CRC further reduced, and there was no significant improvement concerning the computational cost. When it comes to creating deeper convolutional layers instead of using CSEN designs, such as the modified ReconNet, the results presented in Table \ref{tab:performance} shows us that compact CSEN structures are indeed preferable to achieve superior classification performances compared to deeper networks.

\begin{table}[h!]
\begin{center}
\caption{The overall (cumulative) confusion matrix of the proposed recognition scheme.}
\begin{tabular}{|c|l|l|l|l|l|}
\hline
\multicolumn{1}{|l|}{{\color[HTML]{000000} \textbf{CSEN2}}} & \multicolumn{5}{c|}{{\color[HTML]{000000} \textbf{Predicted}}}                                                                                                       \\ \hline
\multicolumn{1}{|l|}{{\color[HTML]{000000} }}               & {\color[HTML]{FF0000} \textbf{}} & {\color[HTML]{000000} Bacterial} & {\color[HTML]{000000} Viral} & {\color[HTML]{000000} Normal} & {\color[HTML]{000000} Covid-19} \\ \hline
{\color[HTML]{000000} }                                     & {\color[HTML]{000000} Bacterial} & {\color[HTML]{000000} 1818}      & {\color[HTML]{000000} 636}   & {\color[HTML]{000000} 180}    & {\color[HTML]{000000} 126}      \\ \cline{2-6} 
{\color[HTML]{000000} }                                     & {\color[HTML]{000000} Viral}     & {\color[HTML]{000000} 338}       & {\color[HTML]{000000} 959}   & {\color[HTML]{000000} 127}    & {\color[HTML]{000000} 61}       \\ \cline{2-6} 
{\color[HTML]{000000} }                                     & {\color[HTML]{000000} Normal}    & {\color[HTML]{000000} 15}        & {\color[HTML]{000000} 71}    & {\color[HTML]{000000} 1428}   & {\color[HTML]{000000} 65}       \\ \cline{2-6} 
\multirow{-4}{*}{{\color[HTML]{000000} \textbf{Real}}}      & {\color[HTML]{000000} Covid-19}  & {\color[HTML]{000000} 0}         & {\color[HTML]{000000} 3}     & {\color[HTML]{000000} 4}      & {\color[HTML]{000000} 455}      \\ \hline
\end{tabular}
\label{tab:Confusion Matrix}
\end{center}
\end{table}

Finally, Table \ref{tab:Confusion Matrix} presents the overall (cumulative) confusion matrix of the proposed CSEN-based Covid-19 recognition approach  over the new QaTa-Cov19 Dataset. The most critical mis-classifications are the false-positives, that is, the mis-classified Covid-19 X-ray images. The confusion matrix shows that the proposed approach has mis-classified 7 Covid-19 images (out of 462). The 3 out of 7 misclassifications are still in “Viral Pneumonia” category, which can be an expected confusion due to the viral nature of Covid-19. However, the other four cases are mis-classified as “Normal” which is indeed a severe clinical misdiagnosis. A close look to these false-negatives in Fig. \ref{fig:FN} reveals the fact that they are indeed very similar to normal images where typical Covid-19 patterns are hardly visible even by an expert's naked eye. It is possible that these images come from the patients who were in the very early stages of Covid-19.


\section{Conclusions}
\label{conclusion}
The commonly used methods in Covid-19 diagnosis, namely Reverse Transcription-Polymerase Chain Reaction and Computed Tomogrophy have certain limitations and drawbacks such as long processing times and unacceptably high mis-diagnosis rates. These drawbacks are also shared by most of the recent works in the literature based on deep learning due to the data scarcity from the Covid-19 cases. Although Deep Learning based recognition techniques are dominant in Computer Vision where they achieved state-of-the-art performance, their performance degrades fast due to data scarcity, which is the reality in this problem at hand. This study aims to address such limitations by proposing a robust and highly accurate Covid-19 recognition approach directly from raw X-ray images without any pre- or post-processing. The proposed approach is based on the CSEN that can be seen as a bridge between Deep Learning models and representation-based methods. CSEN uses both a dictionary and a set of training samples to train direct map from the query samples to the sparse support set of representation coefficients. With this unique ability and having the advantage of a compact network, the proposed CSEN-based Covid-19 recognition systems surpass the competing methods and achieve over $98\%$ sensitivity and over $95\%$ specificity. Furthermore, they yield the most computationally efficient scheme in terms of speed and memory. Finally, the largest dataset of X-ray images, QaTa-Cov19 will be released along with this study as a benchmark dataset in this domain. This will, henceforth, accelerate the research efforts globally and support the fight against Covid-19 worldwide.

\ifCLASSOPTIONcaptionsoff
  \newpage
\fi



%
{\small
\bibliographystyle{IEEEtran}
\bibliography{egbib}
}


%








\end{document}